\documentclass[a4paper, 11pt, twocolumn, accepted=2024-05-04]{quantumarticle}
\pdfoutput=1

\usepackage[utf8]{inputenc}
\usepackage[english]{babel}
\usepackage[T1]{fontenc}
\usepackage{amsmath, amssymb}
\usepackage{bm}
\usepackage{dsfont}
\usepackage[dvipsnames]{xcolor}
\usepackage{physics}

\usepackage{hyperref}
\usepackage[numbers, sort&compress]{natbib}

\newcommand{\gibbs}{\mathrm{G}}
\newcommand{\mf}{\mathrm{MF}}
\newcommand{\kBT}{k_\mathrm{B} T}
\newcommand{\larmor}{\omega_\mathrm{L}}
\newcommand{\ext}{\mathrm{ext}}
\newcommand{\qu}{\mathrm{qu}}
\newcommand{\cl}{\mathrm{cl}}
\newcommand{\sys}{\mathrm{S}}
\newcommand{\bath}{\mathrm{B}}
\newcommand{\inter}{\mathrm{int}}
\newcommand{\sysbath}{\mathrm{SB}}
\newcommand{\rc}{\mathrm{RC}}
\newcommand{\counter}{\mathrm{C}}
\newcommand{\eff}{\mathrm{eff}}

\def\exeter{Department of Physics and Astronomy, University of Exeter, Stocker Road, Exeter EX4 4QL, United Kingdom.}
\def\oxford{Department of Engineering Science, University of Oxford, Parks Road, Oxford OX1 3PJ, United Kingdom.}
\def\glasgow{School of Physics and Astronomy, University of Glasgow, Glasgow, G12 8QQ, United Kingdom.}
\def\macquarie{Department of Physics and Astronomy, Macquarie University, 2109 NSW, Australia.}
\def\potsdam{Institut f\"{u}r Physik und Astronomie, University of Potsdam, 14476 Potsdam, Germany.}

\begin{document}

\title{Enhanced entanglement in multi-bath spin-boson models}

\author{Charlie R. Hogg}
\email[she/her/hers ]{c.r.hogg@exeter.ac.uk}
\affiliation{\exeter}
\orcid{0000-0001-9287-6541}

\author{Federico Cerisola}
\affiliation{\exeter}
\affiliation{\oxford}
\orcid{0000-0003-2961-739X}

\author{James D. Cresser}
\affiliation{\exeter}
\affiliation{\glasgow}
\affiliation{\macquarie}

\author{Simon A. R. Horsley}
\affiliation{\exeter}

\author{Janet Anders}
\affiliation{\exeter}
\affiliation{\potsdam}
\orcid{0000-0002-9791-0363}

\begin{abstract}

The spin-boson model usually considers a spin coupled to a single bosonic bath. However, some physical situations require coupling of the spin to multiple environments. For example, spins interacting with phonons in three-dimensional magnetic materials. Here, we consider a spin coupled isotropically to three independent baths. We show that coupling to multiple baths can significantly increase entanglement between the spin and its environment at zero temperature. The effect of this is to reduce the spin's expectation values in the mean force equilibrium state. In contrast, the classical three-bath spin equilibrium state turns out to be entirely independent of the environmental coupling. These results reveal purely quantum effects that can arise from multi-bath couplings, with potential applications in a wide range of settings, such as magnetic materials.

\end{abstract}

\maketitle

\section{Introduction}

Entanglement is a uniquely quantum resource that has proved fundamental to the development of quantum information, such as quantum computing and quantum key distribution \cite{nielsen_book}. Understanding how entanglement arises in microscopic systems is, therefore, vital for the development of quantum technologies \cite{horodecki2009}. There is an ever-increasing body of work that investigates the role entanglement plays in thermodynamics \cite{goold2016} and whether it can be exploited as a resource \cite{huang2013, hewgill2018, bresque2021}. However, identifying what unique effects and potential advantages entanglement and other uniquely quantum effects can have on thermodynamic processes in general settings still remains one of the key questions of the field of quantum thermodynamics \cite{goold2016}.

In the study of open quantum systems, the spin-boson (SB) model has long served as the standard model for the dynamics of a two-level system \cite{boudjada2014}, analogous to the Caldeira-Leggett model for quantum Brownian motion \cite{caldeira1983}. It describes a two-level quantum system - the spin - interacting with an environment modelled as a bosonic bath of harmonic oscillators \cite{breuer_book}. It has proved highly successful in a variety of settings, from modelling spontaneous emission in two-level systems \cite{lemmer2018, lambert2019} to probing coherences in double quantum dots \cite{guarnieri2018, purkayastha2020}. In particular, the presence of entanglement between the spin and its environment is well-documented in single-bath SB systems \cite{costi2003, lambert2005, kopp2007, amico2008, bera2014, wasesatama2022}.

For many practical applications, however, we wish to consider a spin subjected to multiple sources of noise. Indeed, from recovering classical equations of magnetisation dynamics \cite{anders2022} to modelling multiple vibrational interactions in single-molecule junctions \cite{thomas2019}, multi-bath couplings naturally arise in a variety of settings. Therefore, it is highly relevant to understand the unique effects that arise due to the presence of multiple baths. However, scaling up to multiple baths is not trivial since the contributions from each bath are generally non-additive \cite{giusteri2017, mitchison2018, kolodynski2018, mcconnell2019, gribben2022}. Indeed, assuming additivity has been shown to lead to unphysical results. The authors of \cite{mcconnell2019} explain this practically in the context of electron counting statistics. There, they show how, under such an assumption, currents through a double quantum dot coupled to multiple environments can become non-zero, even when there is no lead bias. Furthermore, the effect of multiple baths on the spin-boson model quantum phase transition has been explored \cite{castroneto2003, novais2005, guo2012, otsuki2013, kohler2013, bruognolo2014, cai2019, weber2023}. In particular, the purely-quantum effect of coupling via non-commuting operators has been found to lead to a ‘frustration of decoherence’ \cite{castroneto2003, novais2005}, which prevents the localisation of the spin into a common bath eigenstate. Moreover, the heat currents through a two-level system arising from coupling to multiple environments held at different temperatures \cite{antosztrikacs2021twobath} can be altered when the operators that couple the spin to the environment do not commute \cite{kato2016, duan2020, antosztrikacs2022transport}. Nevertheless, given the increased complexity of multi-bath couplings, many questions still remain about how quantum effects manifest in these models \cite{amico2008}. 

To explore this, one choice is to consider a spin that couples isotropically to three baths. This is natural because, as an angular momentum operator, spin has three components and each of them can couple to a bath. Three-bath spin-boson models are of direct relevance for the modelling of magnetic materials, where only three-bath models recover the well-known Landau-Lifschitz-Gilbert (LLG) equation \cite{anders2022}. Additionally, the presence of three independent bosonic baths is required to described phonons within the underpinning lattices of bulk materials \cite{nemati2022}.

In this paper, we thus consider an isotropic three-bath SB model, where the spin is coupled arbitrarily strongly to the environment. We focus here on the spin equilibrium states: understanding the static properties of such systems is vital, for example, in magnetic modelling \cite{evans2014, barker2019, usov2020}. Specifically, we demonstrate how the entanglement between a spin and its environment at zero temperature can be enhanced in the SB model by coupling to multiple baths. This, in turn, reduces the spin expectation values. We also highlight the inherently quantum nature of this effect by comparison to the classical three-bath model and its spin equilibrium states.

\section{Three-bath spin-boson model} \label{sec:model}

\begin{figure}
\includegraphics{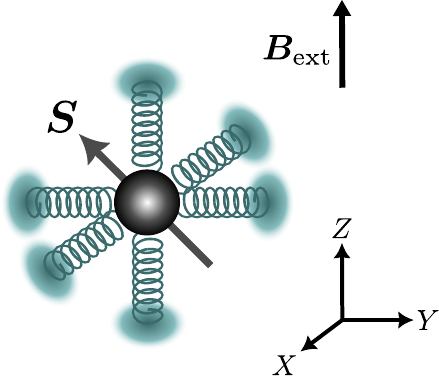}
    \centering
    \caption{\textbf{Three-bath spin-boson model}. A single spin coupled to three bosonic heat baths via the $S_x$, $S_y$ and $S_z$ components of the spin vector (classical) or vector of operators (quantum) $\bm{S}$ \cite{anders2022}. Here, the external field $\bm{B}_\ext$ is aligned in $z$.}
\label{fig:coupling}
\end{figure}

In the following, we will consider a single spin coupled to three bosonic heat baths \cite{anders2022}, as illustrated in FIG. ~\ref{fig:coupling}. Using an open system model, we write the total Hamiltonian
\begin{equation} \label{eq:total_hamiltonian}
    H = H_\sys + H_\bath + H_\inter,
\end{equation}
where $H_\sys$ is the Hamiltonian for the system (spin), $H_\bath$ the Hamiltonian for the baths, and $H_\inter$ the Hamiltonian characterising the interaction between the system and baths \cite{breuer_book}.
For a spin in an external magnetic field $\bm{B}_\ext$, we have
\begin{equation} \label{eq:sys_hamiltonian}
    H_\sys = - \gamma \bm{B}_\ext \cdot \bm{S},
\end{equation}
where $\gamma$ is the gyromagnetic ratio of the spin, and $\bm{S} = (S_x, S_y, S_z)$ is either a vector of spin components for the classical case or vector of spin operators for the quantum case, both with spin length $S_0 = \hbar / 2$. For the following, we will align $\bm{B}_\ext$ along $z$. The Hamiltonian for the baths is given by \cite{huttner1992}
\begin{equation} \label{eq:bath_hamiltonian}
    H_\bath = \frac{1}{2} \int_0^\infty \dd\omega \left( \left( \bm{\mathit{\Pi}}_\omega \right) ^2 + \omega^2 \left( \bm{X}_\omega \right)^2 \right)
\end{equation}
where $\bm{\mathit{\Pi}}_\omega = (\mathit{\Pi}_{\omega, x}, \mathit{\Pi}_{\omega, y}, \mathit{\Pi}_{\omega, z})$ and $\bm{X}_\omega = (X_{\omega}, Y_{\omega}, Z_{\omega})$ are the vectors of three-bath momentum and position operators of the oscillator mode at frequency $\omega$, respectively. Again, these will either be components of a vector classically or of a vector of operators in the quantum case obeying the canonical commutation relations, e.g. $[X_\omega, \mathit{\Pi}_{\omega', x}] = i \hbar \delta(\omega - \omega')$. In what follows, we will be considering \textit{isotropic} bath couplings, so the three-bath interaction Hamiltonian takes the form \cite{anders2022}
\begin{align} \label{eq:int_hamiltonian}
        H_\inter = &- \bm{S} \cdot \int_0^\infty \dd\omega \, c_\omega \bm{X}_\omega \nonumber \\
        = &- S_x \int_0^\infty \dd\omega \, c_\omega X_\omega - S_y \int_0^\infty \dd\omega \, c_\omega Y_\omega \nonumber \\
        &- S_z \int_0^\infty \dd\omega \, c_\omega Z_\omega,
\end{align}
with $c_\omega$ a scalar that sets the coupling strength between the spin and bath mode at frequency $\omega$. For a study of anisotropic couplings, where $c_\omega$ is a tensor, see \cite{hartmann2023}. The spectral density $J(\omega)$ is related to the coupling function through $J(\omega) = c_\omega^2/2\omega$ \cite{cerisola2024}. Setting $c_\omega$ to be of Ohmic form (i.e. $J(\omega) \propto \omega$ \cite{breuer_book}) not only allows for the recovery of the LLG equation \cite{anders2022}, but also has justification from condensed matter physics. Specifically, the resulting spectral density is equivalent to using a Debye density of states, which is commonly used to model three-dimensional materials \cite{nemati2022}. However, recent experiments have revealed the presence of memory effects in magnetic materials at ultrafast timescales \cite{neeraj2021}, which requires one to go beyond a linear spectral density to regimes where these effects naturally arise \cite{anders2022}. An ideal choice of spectral density to model these strong-coupling effects is a Lorentzian \cite{anders2022}, which we will employ here.

One of the key questions in magnetism is how to model the equilibrium properties of magnetic materials \cite{evans2014, barker2019, usov2020}. The equilibrium state of the system interacting with a bath at temperature $T$ is routinely given as the Gibbs state \cite{trushechkin2022}
\begin{equation} \label{eq:gibbs}
    \tau_\gibbs = \frac{1}{\mathcal{Z}_\sys} e^{- \beta H_\sys},
\end{equation}
where $\beta = 1/\kBT$, $k_\mathrm{B}$ is Boltzmann's constant, and $\mathcal{Z}_\sys = \tr [ e^{- \beta H_s} ]$ is the partition function of the system. Note here how the state only depends on the temperature of the environment and not on the strength or form of the interactions \cite{trushechkin2022}. However, on the quantum scale, these interactions naturally become relevant, and so the equilibrium state of the system often differs from the canonical Gibbs state \cite{trushechkin2022}. In such regimes, one can instead employ the `mean force' (MF) state \cite{binder_book_hmf, thingna2012, subasi2012, cresser2021, merkli2022_1, merkli2022_2, trushechkin2022, cerisola2024}, which has been used throughout biology and chemistry to capture the effect of environmental interactions at the smallest of scales \cite{kirkwood1935, lu2021} 
\begin{equation} \label{eq:mfgs}
    \tau_\mf = \tr_\bath \left[ \frac{1}{\mathcal{Z}} e^{- \beta H} \right],
\end{equation}
where $\mathcal{Z} = \tr [ e^{-\beta H} ]$ is the total partition function. Whilst computing the classical MF state at strong coupling is tractable, tracing out the environmental degrees of freedom in the quantum case is much tricker \cite{cerisola2024}. Indeed, analytical expressions only exist for particular regimes and for a single bath \cite{cresser2021}. It remains an open question of how to extend these results to general coupling strengths and multiple baths. One way to solve this issue numerically is by using the reaction coordinate (RC) mapping technique.

\section{Reaction coordinate mapping} \label{sec:rc_mapping}
\begin{figure}
    \includegraphics{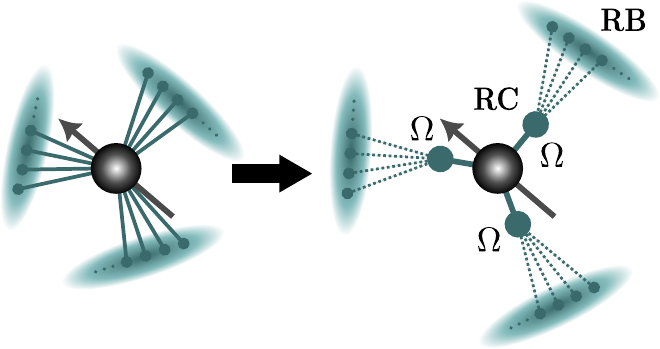}
    \centering
    \caption{\textbf{Three-bath reaction coordinate mapping.} On the left is the original system: a spin coupled strongly to three bosonic heat baths. On the right is the transformed system with three modes at frequency $\Omega$, the reaction coordinates (RCs), extracted from each of the bosonic heat baths, leaving the residual baths (RBs). By appropriately choosing forms of the spectral density defining the coupling between the spin and each RC, and the RCs and RBs, the systems can be shown to generate the same dynamical propagators \cite{garg1985}.}
    \label{fig:rc_mapping}
\end{figure}
This method was first introduced by Garg \textit{et al.} \cite{garg1985} in 1985 to probe strongly-coupled biological and chemical systems. Since then, it has been employed to study a variety of physical settings \cite{ilessmith2014, ilessmith2016, strasberg2016, binder_book_rc, correa2019, antosztrikacs2021twobath, antosztrikacs2021, antosztrikacs2022transport}. This technique involves extracting a mode (the RC) from the bath and coupling the subsystem exclusively to it, and then coupling the RC weakly to a residual bath (RB). As long as the parameters and spectral densities are chosen appropriately, this transformation can be shown to produce identical system dynamics \cite{garg1985}. The RC method has been successfully used to study mean force corrections in the standard SB model \cite{cerisola2024}. Here, we extend the RC method introduced in \cite{garg1985} to three baths, in a similar manner to the two-bath model of \cite{antosztrikacs2022transport}. This is shown Fig.~\ref{fig:rc_mapping}, where the spin now couples to three RCs. Firstly, we define a transformed Hamiltonian as follows
\begin{equation}
\label{eq:tr_hamiltonian}
    H' = H_\sys' + H_\bath' + H_\inter' + H_\counter'.
\end{equation}
Here, $H_\sys'$ is the new, augmented system Hamiltonian which includes the spin, RCs, and the interaction between them
\begin{equation}
\label{eq:tr_sys_hamiltonian}
    H_\sys' = - \gamma \bm{B}_\ext \cdot \bm{S} + \frac{1}{2} \left( \bm{p}^2 + \Omega^2 \bm{x}^2 \right) + \lambda \bm{S} \cdot \bm{x},
\end{equation}
where $\bm{p}$, $\Omega$, and $\bm{x}$ are the momentum, frequency, and position of the three RCs, respectively, and $\lambda$ is the interaction strength between the spin and each RC. Note here that $\lambda$ and $\Omega$ are the same for each bath, given that in Eq.~\eqref{eq:int_hamiltonian} we have considered the same spectral density for each bath. The Hamiltonian $H_\bath'$ is now for the residual baths
\begin{equation}
\label{eq:tr_bath_hamiltonian}
    H_\bath' = \frac{1}{2} \int_0^\infty \dd\omega \left( \left( \bm{\mathit{\Pi}}_\omega' \right)^2 + \omega^2 \left( \bm{X}_\omega' \right)^2 \right),
\end{equation}
where $\bm{\mathit{\Pi}}_\omega'$ and $\bm{X}_\omega'$ are the three-bath momentum and position operators for the RB modes at frequency $\omega$, respectively. We define the interaction Hamiltonian $H_\inter'$ between the augmented system and the RBs as follows
\begin{equation}
\label{eq:tr_int_hamiltonian}
    H_\inter' = \bm{x} \cdot \int_0^\infty \dd\omega \, c_\omega^\rc \bm{X}'_\omega,
\end{equation}
where $c_\omega^\rc$ is a new coupling function for the interaction between the augmented spin-RC system and the RBs at frequency $\omega$. Finally, the counter term for the RCs is given by
\begin{equation}
\label{eq:counter_term}
    H_\counter' = \bm{x}^2 \, \int_0^\infty \dd\omega \, \frac{ \left( c_\omega^\rc \right)^2}{2 \omega^2}.
\end{equation}
This counter term originates from the interaction of the RC harmonic oscillators with the residual baths \cite{breuer_book}. It is worth noting that no such counter term appears in Eq.~\eqref{eq:total_hamiltonian}. In fact, for spin~-~1/2, the counter term would be proportional to the identity, and so it can be safely ignored. We now set each RC-RB spectral density $J^\rc(\omega) = (c_\omega^\rc)^2 / 2 \omega$ to be Ohmic
\begin{equation}\label{eq:tr_spectral_density}
    J^\rc(\omega) = \frac{1}{\pi} \Gamma \omega \frac{\Lambda^2}{\omega^2 + \Lambda^2},
\end{equation}
where $\Gamma$ is the dissipation strength and $\Lambda$ the cutoff frequency, which is taken to be larger than all other relevant frequencies. The propagators for the original and transformed systems are identical when the spectral density for Eq.~\eqref{eq:int_hamiltonian} takes a Lorentzian form \cite{garg1985}
\begin{equation}
\label{eq:original_spectral_density}
    J(\omega) = \frac{1}{\pi} \frac{\lambda^2 \Gamma \omega}{(\Omega^2 - \omega^2)^2 + \Gamma^2 \omega^2},
\end{equation}
with resonant frequency $\Omega$ and peak width $\Gamma$. The closed form of the mapping here demonstrates the benefit of choosing such spectral densities. Here, we will express the Lorentzian parameters as multiples of the Larmor frequency $\larmor = \gamma B_\ext$. As in \cite{anders2022}, we will write the coupling amplitude $\lambda^2 = \larmor^2 \alpha / S_0$, introducing a new parameter $\alpha$ with units of frequency. This allows us to express the spin-RC coupling strength as
\begin{equation} \label{eq:lambda}
    \lambda = \larmor \sqrt{\frac{2 \alpha}{\hbar}}.
\end{equation}
We now have an expression for the RC Hamiltonian \eqref{eq:tr_hamiltonian}, which has been shown to generate the same propagators as the untransformed Hamiltonian \eqref{eq:total_hamiltonian} \cite{garg1985}. For small enough $\Gamma$, the augmented spin and RC system is weakly coupled to the RBs. The quantum mean force (QMF) state can then be obtained by taking the Gibbs state of the augmented system and tracing out all three RCs \cite{binder_book_rc}
\begin{equation} \label{eq:mfgs_rc}
    \tau_\mf^\qu = \tr_\rc \left[ \frac{e^{- \beta H_\sys'}}{\mathcal{Z_\sys}'} \right],
\end{equation}
where $\mathcal{Z}_\sys' = \tr [ e^{\beta H_\sys'} ]$ is the partition function for the spin-RC system. Note that the assumption of small $\Gamma$ imposes no restriction on the strength of the original coupling between the spin and baths, given that $\lambda$ can be arbitrarily large \cite{cerisola2024}. We now have all the tools required to calculate the classical and quantum equilibrium spin expectation values $s_{x, y, z} = \tr[\tau_\mf S_{x, y, z}]/ S_0$ for the three-bath model.

\section{Results and discussion} \label{sec:results}

\subsection{Entanglement in the three-bath model} \label{sec:three_vs_single_bath}
\begin{figure*}
    \centering
    \includegraphics{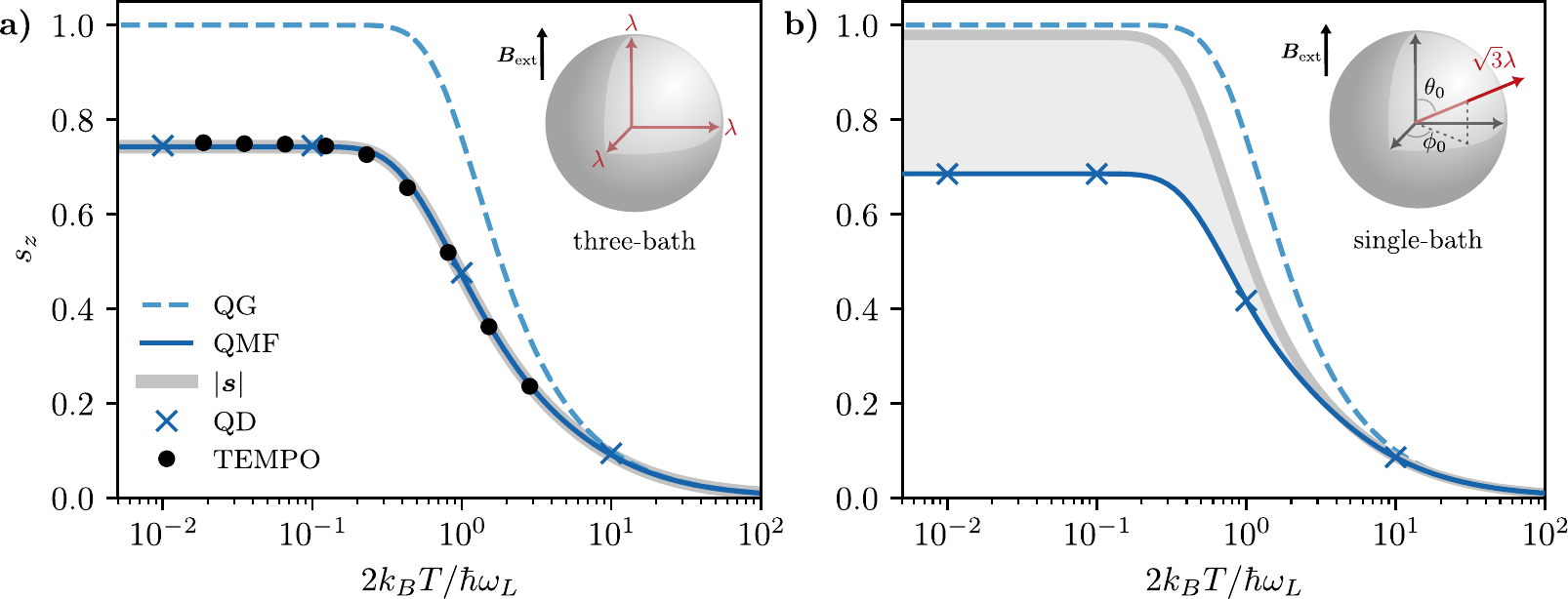}
    \caption{\textbf{Impact of multi-bath coupling on spin-boson equilibrium states.} Panel \textbf{a)} shows the normalised equilibrium expectation $s_z = \langle S_z \rangle / S_0$ for a spin coupled isotropically to three baths $X$, $Y$ and $Z$ via the $S_x$, $S_y$ and $S_z$ operators, respectively - see Eq.~\eqref{eq:int_hamiltonian}, whilst \textbf{b)} shows $s_z$ for a spin coupled to a single bath $R$ through the  $S_R = (S_x + S_y + S_z)/\sqrt{3}$ operator - see Eq.~\eqref{eq:int_hamiltonian_1D}. The solid blue line is $s_z$ for the quantum mean force (QMF) state, see Eq.~\eqref{eq:mfgs_rc}. The key difference between \textbf{a)} and \textbf{b)} is that the spin components orthogonal to $\bm{B}_\ext$, $s_{x, y}$, are only non-zero for the single-bath model, as seen from the difference (shaded) between $|\bm{s}|$ (grey line) and $s_z$. At $T=0$~K, this indicates significantly more entanglement between the spin and its environment in the three-bath model. To verify the numerical accuracy of the QMF state, we also plot the quantum dynamical steady states from the RC master equation (QD, blue crosses) \cite{ilessmith2014} and TEMPO (black dots) \cite{OQuPy, strathearn2018, gribben2022}. The spin length is $S_0 = \hbar / 2$. For the QMF, the parameters are $\omega_0 = 2 \larmor$, $\Gamma = 0.6 \larmor$, $\alpha = 10 \larmor$, and the spin-RC coupling $\lambda$ is given in \eqref{eq:lambda}. For the QD, we further set $\Lambda = 10^{10} \omega_{\mathrm{L}}$. The quantum Gibbs state (QG, dashed line, Eq.~\eqref{eq:gibbs}) is shown as a reference.}
    \label{fig:sz_T_1D_3D}
\end{figure*}
Here, we demonstrate how the presence of multiple baths can have drastic effects on the equilibrium properties of the spin. Using the RC mapping method, we evaluate the expectation values $s_{x, y, z}$ for the QMF for the three-bath SB model \eqref{eq:total_hamiltonian} - see Eq.~\eqref{eq:mfgs_rc}. Fig.~\ref{fig:sz_T_1D_3D}a) shows $s_z$ as a function of temperature $T$ (solid blue). It is immediately obvious that the QMF $s_z$ at $T=0$~K starts at a reduced level, compared to the quantum Gibbs state (QG, dashed blue), where the coupling between spin and baths is negligible. For both the QMF and QG, the two spin components $s_{x,y}$ orthogonal to the external field $\bm{B}_\ext$ are zero for all temperatures. For the QMF, this is highlighted by the match of the $s_z$ expectation value with $|\bm{s}| = \sqrt{s_x^ 2 + s_y^2 + s_z^2}$ (grey line).

The three-bath interaction Hamiltonian \eqref{eq:int_hamiltonian} couples the spin to three independent baths, as shown in the inset of Fig.~\ref{fig:sz_T_1D_3D}a). We now wish to contrast this with the commonly studied case in which the spin couples to a single bath. Of course, it is worth noting that the three-bath and single-bath models will always be fundamentally different for any choice of single-bath coupling direction. However, as we argue in the following, there is a natural choice of single-bath coupling direction that makes such a comparison fair. We proceed by coupling each spin component in \eqref{eq:int_hamiltonian} to the \textit{same} bath with position operator $R_\omega$, i.e. the interaction Hamiltonian becomes
\begin{align}
    H_\inter^{1\mathrm{B}} &= - (S_x + S_y + S_z) \, \int_0^\infty d\omega \, c_\omega R_\omega \nonumber \\ 
    &= - \sqrt{3} \, S_R \, \int_0^\infty d\omega \, c_\omega R_\omega. \label{eq:int_hamiltonian_1D}
\end{align}
Here, $S_R = (S_x + S_y + S_z)/\sqrt{3} = \sin{\theta_0} \cos{\phi_0} \, S_x + \sin{\theta_0} \sin{\phi_0} \, S_y + \cos{\theta_0} \, S_z$ is the spin operator in the direction given by spherical angles $\theta_0 = \arctan{\sqrt{2}}$ and $\phi_0 = \pi/4$. Notice here how $\sqrt{3}$ appears naturally as a normalisation to the system-bath coupling function $c_\omega$. We illustrate this coupling graphically in the inset of Fig.~\ref{fig:sz_T_1D_3D}b). The interaction terms \eqref{eq:int_hamiltonian} and \eqref{eq:int_hamiltonian_1D} permit a fair comparison between the three- and single-bath models in the sense that they do not change the `weight' of the contribution of each spin component operator to the interaction energy, nor the overall strength of the interaction. In fact, in the three-bath case \eqref{eq:int_hamiltonian}, the coupling to the environment must always have both a longitudinal as well as transversal component with respect to the external field. To have a fair comparison, the single-bath case should also have both components and in proportional amount. This fixes the single-bath coupling direction as the one chosen here, see \eqref{eq:int_hamiltonian_1D}. As before, we use the RC mapping method to compute $s_z$ for the single-bath QMF (solid blue), which is plotted as a function of $T$ in Fig.~\ref{fig:sz_T_1D_3D}b).

\medskip

We will now proceed to quantify the zero-temperature entanglement between the spin and baths for both models. Whilst there are many measures of quantum entanglement, e.g. concurrence and logarithmic negativity \cite{nielsen_book}, many are hard to compute \cite{plenio2007}. However, in the limit $T \to 0$~K, the global (spin+bath) Gibbs state becomes pure. 
The reduced state of the spin, $\tau_\mf^\qu$, is then mixed whenever the spin is entangled with the baths. The presence of these correlations can be quantified by the \textit{entanglement entropy} \cite{nielsen_book}
\begin{equation} \label{eq:entanglement_entropy}
    \mathcal{E}(\tau_\mf^\qu) = - \tr \left[ \tau_\mf^\qu \ln \tau_\mf^\qu \right].
\end{equation}

The entanglement-caused mixedness can be further quantified by the spin's purity $\mathcal{P}$. For a \mbox{spin-1/2}, this is~\cite{nielsen_book} 
\begin{equation}
    \mathcal{P}(\tau_\mf^\qu) = \tr \left[ \left( \tau_\mf^\qu \right)^2 \right] = \frac{1}{2} \left( 1 + |\bm{s}|^2 \right).
\end{equation}
The purity is minimal, $\mathcal{P} = 1/2$, when the spin-magnitude is zero, $|\bm{s}| = 0$, i.e. all three spin components must be zero. Here, the entanglement is maximal, $\mathcal{E}_\mathrm{max} = \ln{2}$. On the other hand, the maximal purity is $\mathcal{P} = 1$, which occurs whenever $|\bm{s}| = 1$. Here, the spin state is pure, and hence there is no spin-bath entanglement.

This allows us to link the $T=0$~K value of $|\bm{s}|$ from Fig.~\ref{fig:sz_T_1D_3D} to the entanglement entropy: namely, the lower the value of $|\bm{s}|$, the higher the entanglement entropy between spin and baths. Evaluating Eq.~\eqref{eq:entanglement_entropy} numerically for both models, we find
\begin{align}
    \mathcal{E}_\text{three-bath} &= 0.55\;\mathcal{E}_\mathrm{max}, \label{eq:three_bath_entropy} \\
    \mathcal{E}_\text{single-bath} &= 0.08\;\mathcal{E}_\mathrm{max}, \label{eq:single_bath_entropy}
\end{align}
i.e. the three-bath entanglement entropy is almost a whole order of magnitude greater than the single-bath entanglement entropy. Note that, whilst the entanglement entropies \eqref{eq:three_bath_entropy} and \eqref{eq:single_bath_entropy} are for specific Lorentzian parameters $\omega_0 = 2 \larmor$, $\Gamma = 0.6 \larmor$ and $\alpha = 10 \larmor$, this difference is still observed at other strong coupling values.

Despite a similar reduction in the value of $s_z$ between the single- and three-bath models, see Fig.~\ref{fig:sz_T_1D_3D}, the origin is fundamentally different. Indeed, the spin in the single-bath model `compensates' by increasing $s_{x, y}$. This is clearly demonstrated in Fig.~\ref{fig:sz_T_1D_3D}b) by $|\bm{s}|$ (grey line) being much greater than $s_z$ for the single-bath QMF state at $T \lesssim~\hbar \larmor / 2 k_\mathrm{B}$. We can understand this as follows: in the single-bath case, the primary effect of the interaction with the bath is to align the spin towards the coupling axis \cite{cresser2021, cerisola2024}. This is in stark contrast to the three-bath model where the spin is aligned parallel to $\bm{B}_\ext$, and the sole impact of the environmental coupling is to introduce disorder into the system via entanglement. This is further illustrated by considering the values of $|\bm{s}|^2$ at $T=0$~K: $|\bm{s}|^2 \approx 1$ for the single-bath model, corresponding to a nearly pure state \cite{nielsen_book}. Conversely, $|\bm{s}|^2 \ll 1$ in the three-bath case, meaning that the spin state is much more mixed.

To understand this significant difference between the single and three-bath cases, we can consider the ultrastrong coupling regime. For the single-bath model, we know that the ground state of the spin tends towards an eigenstate of the interaction Hamiltonian in this limit \cite{cresser2021}. The single-bath interaction Hamiltonian \eqref{eq:int_hamiltonian_1D} is a product operator of spin and bath. Hence, its eigenstates are product states. In contrast, for the multi-bath case, since the $S_x$, $S_y$, and $S_z$ coupling operators do not commute, the eigenstates of the interaction Hamiltonian \eqref{eq:int_hamiltonian} do not admit such a decomposition into product states. This is reminiscent of the `frustration of decoherence' that has been studied in the context of phase transitions \cite{novais2005}, and is unique to multi-bath models. In the single-bath case, there is no such competition between baths, which appears to drastically reduce the entanglement between the spin and its environment.

Up until now, we have focused on the zero-temperature case. As temperature increases, we expect entanglement to decrease \cite{anders2008, sadiek2021}, making the low-temperature limit studied here the most relevant case to quantify the effect of multiple baths. It is also worth noting that, whilst we have discussed in detail here the isotropic model, the enhanced entanglement still manifests in the presence of small anisotropies to the system-bath coupling.

The results presented here demonstrate how the presence of multiple dissipation channels in the SB model can have a significant impact on the equilibrium properties of the spin by substantially increasing entanglement with the baths. Hence, a careful assessment is needed to determine if mapping to a single-bath model is sufficient to describe the physics, or whether a multi-bath treatment is needed.

\subsection{Quantum vs. classical models} \label{sec:quantum_vs_classical}

\begin{figure}
    \centering
    \includegraphics{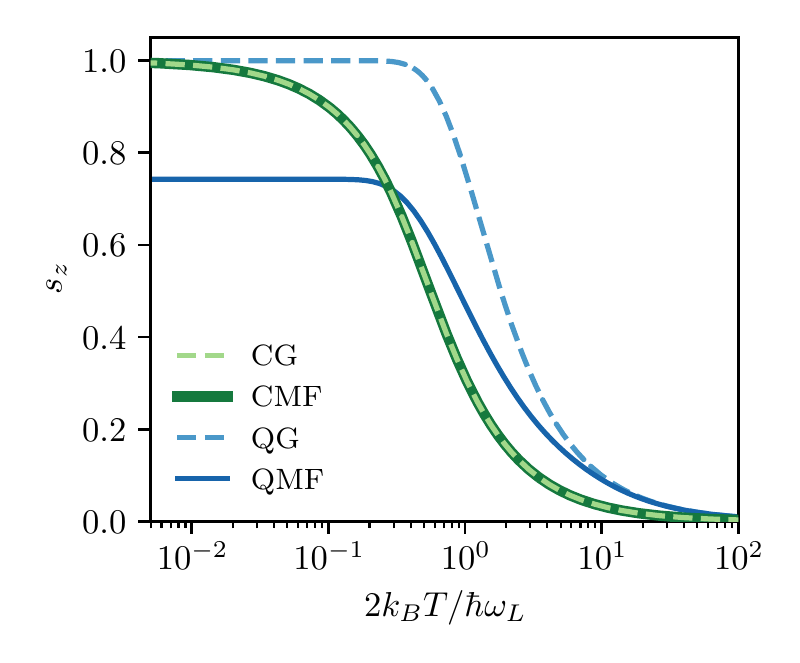}
    \caption{\textbf{Quantum vs. classical isotropic three-bath spin-boson model}. Plotted are the normalised equilibrium spin expectation values $s_z = \langle S_z \rangle / S_0$ over temperature for the quantum mean force (QMF, solid blue, Eq.~\eqref{eq:mfgs_rc}) and classical mean force (CMF, solid green, Eq.~\eqref{eq:cl_mfgs}) states. Whilst the CMF and classical Gibbs state (CG, dashed green, Eq.~\eqref{eq:gibbs}) coincide, this is not the case for the quantum model, where there are stark differences between the QMF and quantum Gibbs (QG, dashed blue, Eq.~\eqref{eq:gibbs}) predictions from entanglement. The spin length is $S_0 = \hbar / 2$, and the parameters for the QMF plots are: $\omega_0 = 2 \larmor$, $\Gamma = 0.6 \larmor$ and $\alpha = 10 \larmor$.}
    \label{fig:sz_T_qu_cl}
\end{figure}

We will now compare the equilibrium states of the quantum and classical versions of the three-bath SB model to highlight the quantum nature of the reduction in $s_z$. Fig.~\ref{fig:sz_T_qu_cl} shows the expectation values $s_z$ for the QMF (solid blue) and the classical mean force state (CMF, solid green, Eq.~\eqref{eq:mfgs}) as a function of temperature $T$.
Remarkably, we find that the environmental corrections to the three-bath QMF, discussed before in Section~\ref{sec:three_vs_single_bath}, vanish entirely for the CMF, which becomes independent of environmental coupling. This is evidenced by the match between expectation values from the CMF and classical Gibbs (CG, dashed green, Eq.~\eqref{eq:gibbs}) states.

To see why no mean force corrections arise classically, one can explicitly evaluate the CMF. For the three-bath isotropic model, the CMF state \eqref{eq:mfgs} takes the form (see Appendix~\ref{sec:three_bath_cmf} for details)
\begin{equation} \label{eq:cl_mfgs}
    \tau_\mf^\cl = \frac{1}{\tilde{\mathcal{Z}}_\sys^\cl} e^{- \beta \left( H_\sys - Q S_0^2 \right)},
\end{equation}
where $\tilde{\mathcal{Z}}_\sys^\cl$ is the partition function that normalises $\tau^\cl_\mf$, and $Q$ the reorganisation energy
\begin{equation}
    Q = \int_0^\infty \dd\omega \frac{J(\omega)}{\omega}.
\end{equation}
Since $S_0^2$ is a constant, the mean-force correction in \eqref{eq:cl_mfgs} is cancelled out entirely by the partition function, and we are left with the standard Gibbs state $\tau_\gibbs^\cl = e^{- \beta H_\sys} / \mathcal{Z}_\sys^\cl$ - see Eq.~\eqref{eq:gibbs}. This is a unique feature of the highly-symmetric three-bath model considered here \cite{hartmann2023}. We highlight that, despite sharing the same symmetry, such a simplification does not occur in the quantum case.
Whilst in the classical isotropic three-bath SB model, the spin equilibrium state is environment-independent, we have shown here that entanglement between the spin and baths plays a key role in the quantum model. We demonstrate that such a correction in the quantum case can have a significant impact on the spin equilibrium properties. Indeed, whilst the quantum and classical spins behave in a qualitatively similar manner in the single-bath model, i.e. by aligning along the coupling axis \cite{cerisola2024}, no such analogue exists for the isotropic three-bath case, where strong-coupling effects are due to entanglement with the environment.

\subsection{Numerical accuracy}

In general, computing the QMF state \eqref{eq:mfgs} proves to be a difficult task, and, as such, we evaluate it numerically using the RC mapping framework. To verify the accuracy of these QMF states (see Figs.~\ref{fig:sz_T_1D_3D} and \ref{fig:sz_T_qu_cl}), we also calculate the steady states from both the RC master equation (QD, blue crosses) \cite{ilessmith2014} and, for the three-bath model, multi-bath TEMPO (TEMPO, black dots) \cite{OQuPy, strathearn2018, gribben2022}. At sufficiently large times, convergence of both these states to the QMF is observed; see Fig.~\ref{fig:sz_T_1D_3D}. The agreement of the TEMPO state is of particular importance as it is entirely independent of the RC mapping and its assumptions.

Aside from verifying our results, this convergence also provides strong numerical evidence that the dynamical steady state of the system is indeed the QMF. Whilst, at least for the single-bath case, this is proven to be true in both the weak \cite{mori2008}, and ultrastrong \cite{trushechkin2022us} limits, there are no such proofs in general for the case of intermediate coupling, especially for multiple baths.

Analytical expressions for the QMF are also limited, although they do exist in particular regimes such as the weak and ultrastrong coupling limits \cite{cresser2021}, and recently an approach has been proposed to tackle the intermediate regime \cite{antosztrikacs2023}. To further establish the numerical accuracy of the QMF, we compute the three-bath weak limit by extending the calculation detailed in \cite{cresser2021} to multiple baths, of which details are given in Appendix~\ref{sec:weak_coupling}. Fig.~\ref{fig:wk} shows $s_z$ calculated using this weak expansion (WK, light grey dots) alongside the zero-temperature limit (WK $T=0$~K, light grey line). As before, we also plot the $s_z$ obtained from the QG (dashed blue line) and QMF (solid blue line) states. We see excellent agreement between the predictions of the QMF and our weak expansion.

\section{Conclusions and outlook} \label{sec:conclusions}
\begin{figure}
    \centering
    \includegraphics{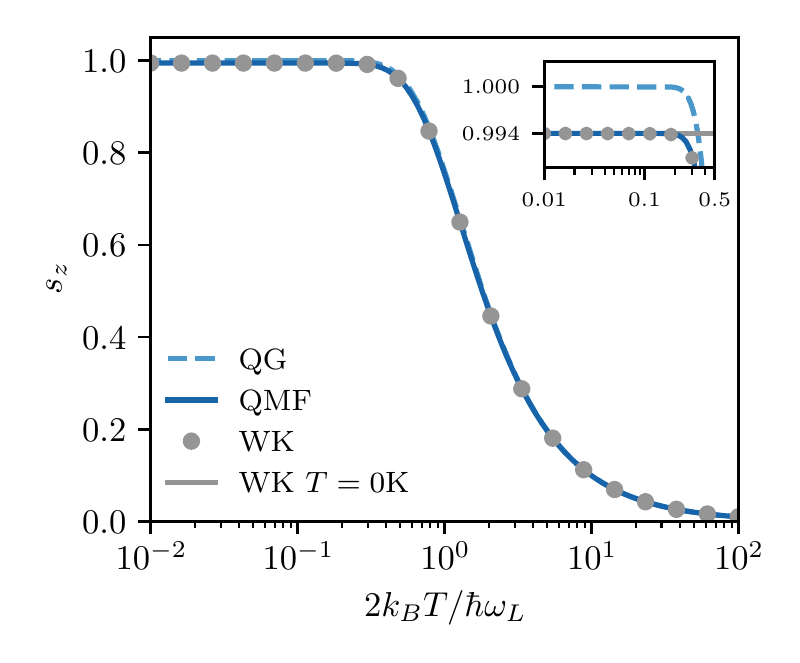}
    \caption{\textbf{Weak coupling limit for a quantum isotropic three-bath spin-boson model}. Plotted are the normalised equilibrium spin expectation values $s_z = \langle S_z \rangle / S_0$ as a function of temperature for a coupling of $\alpha = 0.1 \larmor$. We observe excellent agreement between the predictions of the mean force Gibbs state (QMF, solid blue line), see Eq.~\eqref{eq:mfgs_rc}, and the analytical weak expansion (WK, grey dots) and its zero-temperature limit (WK $T=0$~K, light grey line), see Appendix~\ref{sec:weak_coupling}. The spin length $S_0 = \hbar / 2$, and the parameters for the QMF plots are: $\omega_0 = 2 \larmor$ and $\Gamma = 0.6 \larmor$. The Gibbs state (QG, dashed blue line, Eq.~\eqref{eq:gibbs}) is shown here for reference.}
    \label{fig:wk}
\end{figure}
In this paper, we explored the equilibrium states of a quantum spin coupled isotropically to three bosonic baths. We found that the environmental coupling induces significant entanglement between the spin and baths at zero temperature. We compared these three-bath results to those of a comparable single-bath model, finding substantially more entanglement in the three-bath case. In fact, the strong coupling effects of a single-bath manifest largely as coherences, which have a classical analogue \cite{smith2022, cerisola2024}, whereas there exists no such analogue for entanglement. The enhanced effect of entanglement in the three-bath model leads to marked corrections to the canonical Gibbs state in the form of reduced spin expectation values at zero temperature. This contrasts with the equilibrium state of a classical spin vector coupled isotropically to three baths, for which we proved that no such correction occurs, even for large environmental couplings. Finally, we provided an analytical expression for the three-bath weak coupling state. However, it remains an open question to find a general analytical expression for the spin equilibrium states in other regimes.

The results shown here give insight into the impact of multiple dissipation channels upon spin equilibrium properties of a spin. The direct application is to magnetic modelling, where such multi-bath Hamiltonians arise naturally \cite{anders2022}. Additionally, we expect this work to find use in thermodynamic processes that rely upon the presence of multi-baths, such as thermocurrents \cite{kato2016, duan2020, antosztrikacs2022transport} and quantum heat engines \cite{strasberg2016, newman2017}. Finally, we anticipate that our choice of environmental coupling will add to the interest on non-commuting operators in quantum thermodynamics, for example, in recent work on generalised Gibbs states with non-commuting conserved quantities \cite{yungerhalpern2016, majidy2023}.

\section*{Code availability} 

The code used to produce FIGs.~\ref{fig:sz_T_1D_3D}, \ref{fig:sz_T_qu_cl} and \ref{fig:wk} is available upon reasonable request to CRH, \href{mailto:c.r.hogg@exeter.ac.uk}{c.r.hogg@exeter.ac.uk}.

\section*{Acknowledgements}
We thank Stefano Scali and Pascale Senellart for useful discussions on the subject of this article. We also thank Luis Correa and Nicholas Anto-Sztrikacs for insightful discussions on the RC mapping and Gerald Fux on the usage of TEMPO. CRH is supported by a DTP grant from EPSRC (EP/T518049/1). FC and JA gratefully acknowledge funding from the Foundational Questions Institute Fund (FQXi-IAF19-01). SARH acknowledges funding from the Royal Society and TATA (RPG-2016-186). JA and JC gratefully acknowledge funding from EPSRC (EP/R045577/1). JA thanks the Royal Society for support.

\bibliographystyle{quantum}
\bibliography{references}

\begin{thebibliography}{10}

\bibitem{nielsen_book}
M.~A. Nielsen and I.~L. Chuang.
\newblock ``Quantum computation and quantum information: 10th anniversary
  edition''.
\newblock Cambridge University Press. ~(2010).

\bibitem{horodecki2009}
R.~Horodecki, P.~Horodecki, M.~Horodecki, and K.~Horodecki.
\newblock ``Quantum entanglement''.
\newblock \href{https://dx.doi.org/10.1103/revmodphys.81.865}{Reviews of Modern
  Physics {\bf 81}, 865–942}~(2009).

\bibitem{goold2016}
J.~Goold, M.~Huber, A.~Riera, L.~del Rio, and P.~Skrzypczyk.
\newblock ``The role of quantum information in thermodynamics—a topical
  review''.
\newblock \href{https://dx.doi.org/10.1088/1751-8113/49/14/143001}{Journal of
  Physics A: Mathematical and Theoretical {\bf 49}, 143001}~(2016).

\bibitem{huang2013}
X.~L. Huang, H.~Xu, X.~Y. Niu, and Y.~D. Fu.
\newblock ``A special entangled quantum heat engine based on the two-qubit
  {H}eisenberg {XX} model''.
\newblock \href{https://dx.doi.org/10.1088/0031-8949/88/06/065008}{Physica
  Scripta {\bf 88}, 065008}~(2013).

\bibitem{hewgill2018}
A.~Hewgill, A.~Ferraro, and G.~De~Chiara.
\newblock ``Quantum correlations and thermodynamic performances of two-qubit
  engines with local and common baths''.
\newblock \href{https://dx.doi.org/10.1103/physreva.98.042102}{Physical Review
  A {\bf 98}, 042102}~(2018).

\bibitem{bresque2021}
L.~Bresque, P.~A. Camati, S.~Rogers, K.~Murch, A.~N. Jordan, and
  A.~Auff{\`{e}}ves.
\newblock ``Two-qubit engine fueled by entanglement and local measurements''.
\newblock \href{https://dx.doi.org/10.1103/physrevlett.126.120605}{Physical
  Review Letters {\bf 126}, 120605}~(2021).

\bibitem{boudjada2014}
N.~Boudjada and D.~Segal.
\newblock ``From dissipative dynamics to studies of heat transfer at the
  nanoscale: analysis of the spin-boson model''.
\newblock \href{https://dx.doi.org/10.1021/jp5091685}{The Journal of Physical
  Chemistry A {\bf 118}, 11323–11336}~(2014).

\bibitem{caldeira1983}
A.~O. Caldeira and A.~J. Leggett.
\newblock ``Path integral approach to quantum {B}rownian motion''.
\newblock \href{https://dx.doi.org/10.1016/0378-4371(83)90013-4}{Physica A:
  Statistical Mechanics and its Applications {\bf 121}, 587–616}~(1983).

\bibitem{breuer_book}
H.-P. Breuer and F.~Petruccione.
\newblock ``The theory of open quantum systems''.
\newblock Oxford University Press. ~(2002).

\bibitem{lemmer2018}
A.~Lemmer, C.~Cormick, D.~Tamascelli, T.~Schaetz, S.~F. Huelga, and M.~B.
  Plenio.
\newblock ``A trapped-ion simulator for spin-boson models with structured
  environments''.
\newblock \href{https://dx.doi.org/10.1088/1367-2630/aac87d}{New Journal of
  Physics {\bf 20}, 073002}~(2018).

\bibitem{lambert2019}
N.~Lambert, S.~Ahmed, M.~Cirio, and F.~Nori.
\newblock ``Modelling the ultra-strongly coupled spin-boson model with
  unphysical modes''.
\newblock \href{https://dx.doi.org/10.1038/s41467-019-11656-1}{Nature
  Communications {\bf 10}, 3721}~(2019).

\bibitem{guarnieri2018}
G.~Guarnieri, M.~Kol{\'{a}}{\ifmmode\check{r}\else \v{r}\fi{}}, and R.~Filip.
\newblock ``Steady-state coherences by composite system-bath interactions''.
\newblock \href{https://dx.doi.org/10.1103/physrevlett.121.070401}{Physical
  Review Letters {\bf 121}, 070401}~(2018).

\bibitem{purkayastha2020}
A.~Purkayastha, G.~Guarnieri, M.~T. Mitchison, R.~Filip, and J.~Goold.
\newblock ``Tunable phonon-induced steady-state coherence in a
  double-quantum-dot charge qubit''.
\newblock \href{https://dx.doi.org/10.1038/s41534-020-0256-6}{npj Quantum
  Information {\bf 6}, 27}~(2020).

\bibitem{costi2003}
T.~A. Costi and R.~H. McKenzie.
\newblock ``Entanglement between a qubit and the environment in the spin-boson
  model''.
\newblock \href{https://dx.doi.org/10.1103/physreva.68.034301}{Physical Review
  A {\bf 68}, 034301}~(2003).

\bibitem{lambert2005}
N.~Lambert, C.~Emary, and T.~Brandes.
\newblock ``Entanglement and entropy in a spin-boson quantum phase
  transition''.
\newblock \href{https://dx.doi.org/10.1103/physreva.71.053804}{Physical Review
  A {\bf 71}, 053804}~(2005).

\bibitem{kopp2007}
A.~Kopp and K.~Le~Hur.
\newblock ``Universal and measurable entanglement entropy in the spin-boson
  model''.
\newblock \href{https://dx.doi.org/10.1103/physrevlett.98.220401}{Physical
  Review Letters {\bf 98}, 220401}~(2007).

\bibitem{amico2008}
L.~Amico, R.~Fazio, A.~Osterloh, and V.~Vedral.
\newblock ``Entanglement in many-body systems''.
\newblock \href{https://dx.doi.org/10.1103/revmodphys.80.517}{Reviews of Modern
  Physics {\bf 80}, 517–576}~(2008).

\bibitem{bera2014}
S.~Bera, A.~Nazir, A.~W. Chin, H.~U. Baranger, and S.~Florens.
\newblock ``Generalized multipolaron expansion for the spin-boson model:
  environmental entanglement and the biased two-state system''.
\newblock \href{https://dx.doi.org/10.1103/physrevb.90.075110}{Physical Review
  B {\bf 90}, 075110}~(2014).

\bibitem{wasesatama2022}
V.~A. Wasesatama and J.~S. Kosasih.
\newblock ``Entanglement dynamics of non-linearly coupled spin-boson model
  using hierarchical equation of motion approach''.
\newblock \href{https://dx.doi.org/10.1088/1742-6596/2243/1/012116}{Journal of
  Physics: Conference Series {\bf 2243}, 012116}~(2022).

\bibitem{anders2022}
J.~Anders, C.~R.~J. Sait, and S.~A.~R. Horsley.
\newblock ``Quantum {B}rownian motion for magnets''.
\newblock \href{https://dx.doi.org/10.1088/1367-2630/ac4ef2}{New Journal of
  Physics {\bf 24}, 033020}~(2022).

\bibitem{thomas2019}
J.~O. Thomas, B.~Limburg, J.~K. Sowa, K.~Willick, J.~Baugh, G.~A.~D. Briggs,
  E.~M. Gauger, H.~L. Anderson, and J.~A. Mol.
\newblock ``Understanding resonant charge transport through weakly coupled
  single-molecule junctions''.
\newblock \href{https://dx.doi.org/10.1038/s41467-019-12625-4}{Nature
  Communications {\bf 10}, 4628}~(2019).

\bibitem{giusteri2017}
G.~G. Giusteri, F.~Recrosi, G.~Schaller, and G.~L. Celardo.
\newblock ``Interplay of different environments in open quantum systems:
  breakdown of the additive approximation''.
\newblock \href{https://dx.doi.org/10.1103/physreve.96.012113}{Physical Review
  E {\bf 96}, 012113}~(2017).

\bibitem{mitchison2018}
M.~T. Mitchison and M.~B. Plenio.
\newblock ``Non-additive dissipation in open quantum networks out of
  equilibrium''.
\newblock \href{https://dx.doi.org/10.1088/1367-2630/aa9f70}{New Journal of
  Physics {\bf 20}, 033005}~(2018).

\bibitem{kolodynski2018}
J.~Ko{\l{}}ody{\ifmmode \acute{n}\else \'{n}\fi{}}ski, J.~B. Brask,
  M.~Perarnau-Llobet, and B.~Bylicka.
\newblock ``Adding dynamical generators in quantum master equations''.
\newblock \href{https://dx.doi.org/10.1103/physreva.97.062124}{Physical Review
  A {\bf 97}, 062124}~(2018).

\bibitem{mcconnell2019}
C.~McConnell and A.~Nazir.
\newblock ``Electron counting statistics for non-additive environments''.
\newblock \href{https://dx.doi.org/10.1063/1.5095838}{The Journal of Chemical
  Physics {\bf 151}, 054104}~(2019).

\bibitem{gribben2022}
D.~Gribben, D.~M. Rouse, J.~Iles-Smith, A.~Strathearn, H.~Maguire, P.~Kirton,
  A.~Nazir, E.~M. Gauger, and B.~W. Lovett.
\newblock ``Exact dynamics of nonadditive environments in non-{M}arkovian open
  quantum systems''.
\newblock \href{https://dx.doi.org/10.1103/prxquantum.3.010321}{PRX Quantum
  {\bf 3}, 010321}~(2022).

\bibitem{castroneto2003}
A.~H. Castro~Neto, E.~Novais, L.~Borda, G.~Zar{\'{a}}nd, and I.~Affleck.
\newblock ``Quantum magnetic impurities in magnetically ordered systems''.
\newblock \href{https://dx.doi.org/10.1103/physrevlett.91.096401}{Physical
  Review Letters {\bf 91}, 096401}~(2003).

\bibitem{novais2005}
E.~Novais, A.~H. Castro~Neto, L.~Borda, I.~Affleck, and G.~Zarand.
\newblock ``Frustration of decoherence in open quantum systems''.
\newblock \href{https://dx.doi.org/10.1103/physrevb.72.014417}{Physical Review
  B {\bf 72}, 014417}~(2005).

\bibitem{guo2012}
C.~Guo, A.~Weichselbaum, J.~von Delft, and M.~Vojta.
\newblock ``Critical and strong-coupling phases in one- and two-bath spin-boson
  models''.
\newblock \href{https://dx.doi.org/10.1103/physrevlett.108.160401}{Physical
  Review Letters {\bf 108}, 160401}~(2012).

\bibitem{otsuki2013}
J.~Otsuki.
\newblock ``Spin-boson coupling in continuous-time quantum monte carlo''.
\newblock \href{https://dx.doi.org/10.1103/physrevb.87.125102}{Physical Review
  B {\bf 87}, 125102}~(2013).

\bibitem{kohler2013}
H.~Kohler, A.~Hackl, and S.~Kehrein.
\newblock ``Nonequilibrium dynamics of a system with quantum frustration''.
\newblock \href{https://dx.doi.org/10.1103/physrevb.88.205122}{Physical Review
  B {\bf 88}, 205122}~(2013).

\bibitem{bruognolo2014}
B.~Bruognolo, A.~Weichselbaum, C.~Guo, J.~von Delft, I.~Schneider, and
  M.~Vojta.
\newblock ``Two-bath spin-boson model: phase diagram and critical properties''.
\newblock \href{https://dx.doi.org/10.1103/physrevb.90.245130}{Physical Review
  B {\bf 90}, 245130}~(2014).

\bibitem{cai2019}
A.~Cai and Q.~Si.
\newblock ``Bose-{F}ermi {A}nderson model with {SU}(2) symmetry:
  continuous-time quantum {M}onte {C}arlo study''.
\newblock \href{https://dx.doi.org/10.1103/physrevb.100.014439}{Physical Review
  B {\bf 100}, 014439}~(2019).

\bibitem{weber2023}
M.~Weber and M.~Vojta.
\newblock ``{SU}(2)-symmetric spin-boson model: quantum criticality,
  fixed-point annihilation, and duality''.
\newblock \href{https://dx.doi.org/10.1103/physrevlett.130.186701}{Physical
  Review Letters {\bf 130}, 186701}~(2023).

\bibitem{antosztrikacs2021twobath}
N.~Anto-Sztrikacs and D.~Segal.
\newblock ``Strong coupling effects in quantum thermal transport with the
  reaction coordinate method''.
\newblock \href{https://dx.doi.org/10.1088/1367-2630/ac02df}{New Journal of
  Physics {\bf 23}, 063036}~(2021).

\bibitem{kato2016}
A.~Kato and Y.~Tanimura.
\newblock ``Quantum heat current under non-perturbative and non-{M}arkovian
  conditions: applications to heat machines''.
\newblock \href{https://dx.doi.org/10.1063/1.4971370}{The Journal of Chemical
  Physics {\bf 145}, 224105}~(2016).

\bibitem{duan2020}
C.~Duan, C.-Y. Hsieh, J.~Liu, J.~Wu, and J.~Cao.
\newblock ``Unusual transport properties with noncommutative system–bath
  coupling operators''.
\newblock \href{https://dx.doi.org/10.1021/acs.jpclett.0c00985}{The Journal of
  Physical Chemistry Letters {\bf 11}, 4080–4085}~(2020).

\bibitem{antosztrikacs2022transport}
N.~Anto-Sztrikacs, F.~Ivander, and D.~Segal.
\newblock ``Quantum thermal transport beyond second order with the reaction
  coordinate mapping''.
\newblock \href{https://dx.doi.org/10.1063/5.0091133}{The Journal of Chemical
  Physics {\bf 156}, 214107}~(2022).

\bibitem{nemati2022}
S.~Nemati, C.~Henkel, and J.~Anders.
\newblock ``Coupling function from bath density of states''.
\newblock \href{https://dx.doi.org/10.1209/0295-5075/ac7b42}{Europhysics
  Letters {\bf 139}, 36002}~(2022).

\bibitem{evans2014}
R.~F.~L. Evans, W.~J. Fan, P.~Chureemart, T.~A. Ostler, M.~O.~A. Ellis, and
  R.~W. Chantrell.
\newblock ``Atomistic spin model simulations of magnetic nanomaterials''.
\newblock \href{https://dx.doi.org/10.1088/0953-8984/26/10/103202}{Journal of
  Physics: Condensed Matter {\bf 26}, 103202}~(2014).

\bibitem{barker2019}
J.~Barker and G.~E.~W. Bauer.
\newblock ``Semiquantum thermodynamics of complex ferrimagnets''.
\newblock \href{https://dx.doi.org/10.1103/physrevb.100.140401}{Physical Review
  B {\bf 100}, 140401(R)}~(2019).

\bibitem{usov2020}
N.~A. Usov and O.~N. Serebryakova.
\newblock ``Equilibrium properties of assembly of interacting superparamagnetic
  nanoparticles''.
\newblock \href{https://dx.doi.org/10.1038/s41598-020-70711-w}{Scientific
  Reports {\bf 10}, 13677}~(2020).

\bibitem{huttner1992}
B.~Huttner and S.~M. Barnett.
\newblock ``Quantization of the electromagnetic field in dielectrics''.
\newblock \href{https://dx.doi.org/10.1103/physreva.46.4306}{Physical Review A
  {\bf 46}, 4306–4322}~(1992).

\bibitem{hartmann2023}
F.~Hartmann, S.~Scali, and J.~Anders.
\newblock ``Anisotropic signatures in the spin-boson model''.
\newblock \href{https://dx.doi.org/10.1103/physrevb.108.184402}{Physical Review
  B {\bf 108}, 184402}~(2023).

\bibitem{cerisola2024}
F.~Cerisola, M.~Berritta, S.~Scali, S.~A.~R. Horsley, J.~D. Cresser, and
  J.~Anders.
\newblock ``Quantum-classical correspondence in spin-boson equilibrium states
  at arbitrary coupling''.
\newblock \href{https://dx.doi.org/10.1088/1367-2630/ad4818}{New Journal of
  Physics}~(2024).

\bibitem{neeraj2021}
K.~Neeraj, N.~Awari, S.~Kovalev, D.~Polley, N.~Zhou~Hagstr{\"{o}}m, S.~S. P.~K.
  Arekapudi, A.~Semisalova, K.~Lenz, B.~Green, J.-C. Deinert, I.~Ilyakov,
  M.~Chen, M.~Bawatna, V.~Scalera, M.~d’Aquino, C.~Serpico, O.~Hellwig, J.-E.
  Wegrowe, M.~Gensch, and S.~Bonetti.
\newblock ``Inertial spin dynamics in ferromagnets''.
\newblock \href{https://dx.doi.org/10.1038/s41567-020-01040-y}{Nature Physics
  {\bf 17}, 245–250}~(2021).

\bibitem{trushechkin2022}
A.~S. Trushechkin, M.~Merkli, J.~D. Cresser, and J.~Anders.
\newblock ``Open quantum system dynamics and the mean force {G}ibbs state''.
\newblock \href{https://dx.doi.org/10.1116/5.0073853}{AVS Quantum Science {\bf
  4}, 012301}~(2022).

\bibitem{binder_book_hmf}
H.~J.~D. Miller.
\newblock ``Hamiltonian of mean force for strongly-coupled systems''.
\newblock In F.~Binder, L.~A. Correa, C.~Gogolin, J.~Anders, and G.~Adesso,
  editors, Thermodynamics in the quantum regime.
\newblock \href{https://dx.doi.org/10.1007/978-3-319-99046-0_22}{Page
  531–549}.
\newblock Springer International Publishing, Cham~(2018).

\bibitem{thingna2012}
J.~Thingna, J.-S. Wang, and P.~H{\"{a}}nggi.
\newblock ``Generalized {G}ibbs state with modified {R}edfield solution: exact
  agreement up to second order''.
\newblock \href{https://dx.doi.org/10.1063/1.4718706}{The Journal of Chemical
  Physics {\bf 136}, 194110}~(2012).

\bibitem{subasi2012}
Y.~Suba{\ifmmode \mbox{\c{s}}\else \c{s}\fi{}\ifmmode \imath \else \i \fi{}},
  C.~H. Fleming, J.~M. Taylor, and B.~L. Hu.
\newblock ``Equilibrium states of open quantum systems in the strong coupling
  regime''.
\newblock \href{https://dx.doi.org/10.1103/physreve.86.061132}{Physical Review
  E {\bf 86}, 061132}~(2012).

\bibitem{cresser2021}
J.~D. Cresser and J.~Anders.
\newblock ``Weak and ultrastrong coupling limits of the quantum mean force
  {G}ibbs state''.
\newblock \href{https://dx.doi.org/10.1103/physrevlett.127.250601}{Physical
  Review Letters {\bf 127}, 250601}~(2021).

\bibitem{merkli2022_1}
M.~Merkli.
\newblock ``Dynamics of open quantum systems {I}, oscillation and decay''.
\newblock \href{https://dx.doi.org/10.22331/q-2022-01-03-615}{Quantum {\bf 6},
  615}~(2022).

\bibitem{merkli2022_2}
M.~Merkli.
\newblock ``Dynamics of open quantum systems {II}, {M}arkovian approximation''.
\newblock \href{https://dx.doi.org/10.22331/q-2022-01-03-616}{Quantum {\bf 6},
  616}~(2022).

\bibitem{kirkwood1935}
J.~G. Kirkwood.
\newblock ``Statistical mechanics of fluid mixtures''.
\newblock \href{https://dx.doi.org/10.1063/1.1749657}{The Journal of Chemical
  Physics {\bf 3}, 300–313}~(1935).

\bibitem{lu2021}
C.~Lu, C.~Wu, D.~Ghoreishi, W.~Chen, L.~Wang, W.~Damm, G.~A. Ross, M.~K.
  Dahlgren, E.~Russell, C.~D. Von~Bargen, R.~Abel, R.~A. Friesner, and E.~D.
  Harder.
\newblock ``{OPLS}4: improving force field accuracy on challenging regimes of
  chemical space''.
\newblock \href{https://dx.doi.org/10.1021/acs.jctc.1c00302}{Journal of
  Chemical Theory and Computation {\bf 17}, 4291–4300}~(2021).

\bibitem{garg1985}
A.~Garg, J.~N. Onuchic, and V.~Ambegaokar.
\newblock ``Effect of friction on electron transfer in biomolecules''.
\newblock \href{https://dx.doi.org/10.1063/1.449017}{The Journal of Chemical
  Physics {\bf 83}, 4491–4503}~(1985).

\bibitem{ilessmith2014}
J.~Iles-Smith, N.~Lambert, and A.~Nazir.
\newblock ``Environmental dynamics, correlations, and the emergence of
  noncanonical equilibrium states in open quantum systems''.
\newblock \href{https://dx.doi.org/10.1103/physreva.90.032114}{Physical Review
  A {\bf 90}, 032114}~(2014).

\bibitem{ilessmith2016}
J.~Iles-Smith, A.~G. Dijkstra, N.~Lambert, and A.~Nazir.
\newblock ``Energy transfer in structured and unstructured environments: master
  equations beyond the {B}orn-{M}arkov approximations''.
\newblock \href{https://dx.doi.org/10.1063/1.4940218}{The Journal of Chemical
  Physics {\bf 144}, 044110}~(2016).

\bibitem{strasberg2016}
P.~Strasberg, G.~Schaller, N.~Lambert, and T.~Brandes.
\newblock ``Nonequilibrium thermodynamics in the strong coupling and
  non-{M}arkovian regime based on a reaction coordinate mapping''.
\newblock \href{https://dx.doi.org/10.1088/1367-2630/18/7/073007}{New Journal
  of Physics {\bf 18}, 073007}~(2016).

\bibitem{binder_book_rc}
A.~Nazir and G.~Schaller.
\newblock ``The reaction coordinate mapping in quantum thermodynamics''.
\newblock In F.~Binder, L.~A. Correa, C.~Gogolin, J.~Anders, and G.~Adesso,
  editors, Thermodynamics in the quantum regime.
\newblock \href{https://dx.doi.org/10.1007/978-3-319-99046-0_23}{Pages
  551--577}.
\newblock Springer International Publishing, Cham~(2018).

\bibitem{correa2019}
L.~A. Correa, B.~Xu, B.~Morris, and G.~Adesso.
\newblock ``Pushing the limits of the reaction-coordinate mapping''.
\newblock \href{https://dx.doi.org/10.1063/1.5114690}{The Journal of Chemical
  Physics {\bf 151}, 094107}~(2019).

\bibitem{antosztrikacs2021}
N.~Anto-Sztrikacs and D.~Segal.
\newblock ``Capturing non-{M}arkovian dynamics with the reaction coordinate
  method''.
\newblock \href{https://dx.doi.org/10.1103/physreva.104.052617}{Physical Review
  A {\bf 104}, 052617}~(2021).

\bibitem{OQuPy}
The~TEMPO collaboration.
\newblock ``{OQuPy}: a {P}ython 3 package to efficiently compute
  non-{M}arkovian open quantum systems''~(2020).

\bibitem{strathearn2018}
A.~Strathearn, P.~Kirton, D.~Kilda, J.~Keeling, and B.~W. Lovett.
\newblock ``Efficient non-{M}arkovian quantum dynamics using time-evolving
  matrix product operators''.
\newblock \href{https://dx.doi.org/10.1038/s41467-018-05617-3}{Nature
  Communications {\bf 9}, 3322}~(2018).

\bibitem{plenio2007}
M.~B. Plenio and S.~Virmani.
\newblock ``An introduction to entanglement measures''.
\newblock \href{https://dx.doi.org/10.26421/qic7.1-2-1}{Quantum Information and
  Computation {\bf 7}, 1–51}~(2007).

\bibitem{anders2008}
J.~Anders.
\newblock ``Thermal state entanglement in harmonic lattices''.
\newblock \href{https://dx.doi.org/10.1103/physreva.77.062102}{Physical Review
  A {\bf 77}, 062102}~(2008).

\bibitem{sadiek2021}
G.~Sadiek and S.~Almalki.
\newblock ``Thermal robustness of entanglement in a dissipative two-dimensional
  spin system in an inhomogeneous magnetic field''.
\newblock \href{https://dx.doi.org/10.3390/e23081066}{Entropy {\bf 23},
  1066}~(2021).

\bibitem{mori2008}
T.~Mori and S.~Miyashita.
\newblock ``Dynamics of the density matrix in contact with a thermal bath and
  the quantum master equation''.
\newblock \href{https://dx.doi.org/10.1143/jpsj.77.124005}{Journal of the
  Physical Society of Japan {\bf 77}, 124005}~(2008).

\bibitem{trushechkin2022us}
A.~Trushechkin.
\newblock ``Quantum master equations and steady states for the
  ultrastrong-coupling limit and the strong-decoherence limit''.
\newblock \href{https://dx.doi.org/10.1103/physreva.106.042209}{Physical Review
  A {\bf 106}, 042209}~(2022).

\bibitem{antosztrikacs2023}
N.~Anto-Sztrikacs, A.~Nazir, and D.~Segal.
\newblock ``Effective-{H}amiltonian theory of open quantum systems at strong
  coupling''.
\newblock \href{https://dx.doi.org/10.1103/prxquantum.4.020307}{PRX Quantum
  {\bf 4}, 020307}~(2023).

\bibitem{smith2022}
A.~Smith, K.~Sinha, and C.~Jarzynski.
\newblock ``Quantum coherences and classical inhomogeneities as equivalent
  thermodynamics resources''.
\newblock \href{https://dx.doi.org/10.3390/e24040474}{Entropy {\bf 24},
  474}~(2022).

\bibitem{newman2017}
D.~Newman, F.~Mintert, and A.~Nazir.
\newblock ``Performance of a quantum heat engine at strong reservoir
  coupling''.
\newblock \href{https://dx.doi.org/10.1103/physreve.95.032139}{Physical Review
  E {\bf 95}, 032139}~(2017).

\bibitem{yungerhalpern2016}
N.~Yunger~Halpern, P.~Faist, J.~Oppenheim, and A.~Winter.
\newblock ``Microcanonical and resource-theoretic derivations of the thermal
  state of a quantum system with noncommuting charges''.
\newblock \href{https://dx.doi.org/10.1038/ncomms12051}{Nature Communications
  {\bf 7}, 12051}~(2016).

\bibitem{majidy2023}
S.~Majidy, A.~Lasek, D.~A. Huse, and N.~Yunger~Halpern.
\newblock ``Non-{A}belian symmetry can increase entanglement entropy''.
\newblock \href{https://dx.doi.org/10.1103/physrevb.107.045102}{Physical Review
  B {\bf 107}, 045102}~(2023).

\end{thebibliography}

\onecolumn
\appendix

\section{Isotropic three-bath CMF state} \label{sec:three_bath_cmf}

As in \cite{cerisola2024}, we complete the square on \eqref{eq:total_hamiltonian}, allowing us to write the total Hamiltonian as
\begin{equation} \label{eq:total_hamiltonian_cl_1}
    H = - \gamma \bm{B}_\ext \cdot \bm{S} + \frac{1}{2} \int^\infty_0 \dd\omega \, \left( \left( \bm{\mathit{\Pi}}_\omega \right)^2 + \omega^2 \left( \bm{X}_\omega - \bm{\mathit{\mu}}_\omega \right)^2 \right) - \frac{1}{2} \int^\infty_0 \dd\omega \, \omega^2 \bm{\mathit{\mu}}^2_\omega,
\end{equation}
where
\begin{equation} \label{eq:mu}
    \boldsymbol{\mu}_\omega = \frac{c_\omega}{\omega^2} \bm{S}.
\end{equation}
Given the reorganisation energy $Q = \int_0^\infty \dd\omega \, c_\omega^2 / 2\omega^2$ \cite{cerisola2024}, Eq.~\eqref{eq:total_hamiltonian_cl_1} becomes
\begin{equation} \label{eq:total_hamiltonian_cl_2}
    H = H_\eff + H_\bath,
\end{equation}
with
\begin{align}
    H_\eff &= - \gamma \bm{B}_\ext \bm{S} - Q S_0^2, \label{eq:effective_hamiltonian_1} \\
    H_\bath &= \frac{1}{2} \int_0^\infty \dd\omega \, \left( \left( \bm{\mathit{\Pi}}_\omega \right)^2 + \omega^2 \left( \bm{X}_\omega - \bm{\mathit{\mu}}_\omega \right)^2 \right), \label{eq:bath_hamiltonian_cl}
\end{align}
where we have used the fact that $\bm{S}^2 = S_0^2$ is a constant. The CMF state can be found by tracing out the bath as follows
\begin{equation} \label{eq:CMF_1}
    \tau^\cl_\mf = \frac{1}{\mathcal{Z}^\cl_\sysbath} \prod_\omega \int_{- \infty}^\infty \dd\bm{X}_\omega \, \int_{- \infty}^\infty \dd\bm{\mathit{\Pi}}_\omega e^{-\beta (H_\eff + H_\bath)} = \frac{\mathcal{Z}^\cl_\bath}{\mathcal{Z}^\cl_\sysbath} e^{- \beta H_\eff} = \frac{\mathcal{Z}^\cl_\bath}{\mathcal{Z}^\cl_\sysbath} e^{- \beta H_\sys} e^{\beta Q S_0^2},
\end{equation}
where $\mathcal{Z}_\sysbath^\cl$ and $\mathcal{Z}_\bath^\cl$ are the classical total and bath partition functions, respectively. Explicitly evaluating the former gives 
\begin{equation} \label{eq:total_partition_cl}
    \mathcal{Z}^\cl_\sysbath = \int^{S_0}_{- S_0} \dd\bm{S} \, e^{- \beta H_\eff} \prod_\omega \int_{- \infty}^\infty \dd\bm{X}_\omega \, \int_{- \infty}^\infty \dd\bm{\mathit{\Pi}}_\omega \, e^{-\beta H_\bath} = \mathcal{Z}^\cl_\bath \int^{S_0}_{- S_0} \dd\bm{S} \, e^{- \beta H_\eff} = \mathcal{Z}_\bath^\cl \tilde{\mathcal{Z}}_\sys^\cl,
\end{equation}
where $\tilde{\mathcal{Z}}_\sys^\cl$ is the system MF partition function
\begin{equation} \label{eq:system_partition_cl}
    \tilde{\mathcal{Z}}_\sys^\cl = \int^{S_0}_{- S_0} \dd\bm{S} \, e^{- \beta H_\eff} = e^{\beta Q S_0^2} \int^{S_0}_{- S_0} \dd\bm{S} \, e^{- \beta H_\sys} = e^{\beta Q S_0^2} \mathcal{Z}_\sys^\cl,
\end{equation}
where $\mathcal{Z}_\sys^\cl$ is the standard system partition function. Substituting \eqref{eq:system_partition_cl} into \eqref{eq:total_partition_cl}, and then \eqref{eq:total_partition_cl} into \eqref{eq:CMF_1} yields
\begin{equation} \label{eq:CMF_2}
    \tau^\cl_\mf = \frac{1}{\mathcal{Z}^\cl_\sys}e^{- \beta H_\sys},
\end{equation}
which is the standard Gibbs state.

\section{Weak-coupling limit} \label{sec:weak_coupling}

We will write the interaction Hamiltonian \eqref{eq:int_hamiltonian} as
\begin{equation} \label{eq:int_hamiltonian_weakx}
    H_\inter= - \bm{B} \cdot \bm{S},
\end{equation}
where 
\begin{equation}
    \bm{B}= \int_{0}^{\infty} \dd\omega \, c_\omega \bm{X}_\omega.
\end{equation}
The vector operator $\bm{B}$ can be written in terms of its vector components as $\bm{B} = \sum_i B_i \, \mathbf{e}_i$, $i = x, y, z$. Thus the spin is coupled to three baths, which will be assumed to be independent in the sense that $\big[ B_i, B_j \big] = 0$. Each vector operator component can be expressed as
\begin{equation}
    B_i = \int_{0}^{\infty} \dd\omega \, c_\omega X_{\omega, i},
\end{equation}
where
\begin{equation}
    X_{\omega, i} = \sqrt{\frac{\hbar}{2 \omega}} \left( a_{\omega, i} + a_{\omega, i}^\dagger \right).
\end{equation}
For three independent bosonic baths, the annihilation and creation operators $a_{w, i}$ and $a_{\omega', j}^\dagger$ will satisfy the bosonic commutation rules $\big[ a_{\omega, i}, a_{\omega', j}^\dagger \big] = \delta_{ij} \delta(\omega-\omega')$. The bath Hamiltonian will then be
\begin{equation}
    H_\bath = \sum_{i} H_{\bath, i} = \hbar \sum_{i} \int_{0}^{\infty} \dd\omega \, \omega \, a_{\omega, i}^\dagger \, a_{\omega, i},
\end{equation}
with $\left[ H_{\bath, i}, H_{\bath, j} \right]=0$. The Gibbs state of the bath with all components at the common inverse temperature $\beta$ will be
\begin{equation}
    \tau_\bath = \mathcal{Z}_\bath^{-1} e^{- \beta H_\bath} = \mathcal{Z}_\bath^{-1} \prod_i e^{- \beta H_{\bath, i}}.
\end{equation}
We will now suppose that the combined system comes to equilibrium at the same inverse temperature $\beta$ as the initial temperature of the bath. The partition function for the combined system will then be
\begin{equation}
    \mathcal{Z}_\sysbath = \tr_\sysbath \left[e^{- \beta H} \right] = \tr_\sys \left[ \tilde{\rho}_\sys \right],
\end{equation}
where $\tilde{\rho}_\sys = \tr_\bath \big[ e^{- \beta H} \big]$ is the unnormalised reduced state of the system. Expanding $e^{- \beta H}$ to second order in the interaction $H_\inter$, i.e. the Kubo expansion, gives this reduced state as
\begin{equation}
    \tilde{\rho}_\sys^{(2)} = \tr_\bath \left[ e^{-\beta H_0} \left( 1 - \int_{0}^{\beta} \dd\beta' H_\inter(\beta') + \int_{0}^{\beta} \dd\beta' \int_{0}^{\beta'} \dd\beta'' H_\inter(\beta' )H_\inter (\beta'') \right) \right],
\end{equation}
where, for any Schr\"{o}dinger operator $O$, the `thermal interaction' picture operator $O(\beta)$ is given by $O(\beta) = e^{\beta H_0} O e^{- \beta H_0}$, with $H_0 = H_\sys + H_\bath$. We can then write, and noting that $[B_i(\beta), S_j(\beta)] = 0$,
\begin{equation}
    H_\inter(\beta)= - e^{\beta H_\bath} \bm{B} e^{- \beta H_\bath} \cdot e^{\beta H_\sys} \bm{S} e^{- \beta H_\sys} = - \bm{B}(\beta) \cdot \bm{S}(\beta).
\end{equation}
We then find that
\begin{equation}
    \tilde{\rho}_\sys^{(2)} = e^{- \beta H_\sys} \tr_\bath \left[ e^{- \beta H_\bath} \left( 1 + \int_0^\beta \dd\beta' \bm{B}(\beta') \cdot \bm{S}(\beta') + \int_0^\beta \dd\beta' \int_0^{\beta'} \dd\beta'' \bm{B}(\beta') \cdot \bm{S}(\beta') \bm{B} (\beta'') \cdot \bm{S}(\beta'') \right) \right].
\end{equation}
Introducing the partition functions for the bare system and bath, $\tr_\sys \big[ e^{- \beta H_\sys} \big] = \mathcal{Z}_\sys^{(0)}$ and $\tr_\bath \big[e^{- \beta H_\bath} \big] = \mathcal{Z}_\bath$ respectively, using $\tr_\bath \big[ e^{- \beta H_\bath} \bm{B} \big] = 0$, and expanding the scalar products in the double integral we get
\begin{equation}
    \tilde{\rho}_\sys^{(2)} = \mathcal{Z}_\sys^{(0)} \mathcal{Z}_\bath \, \tau_\sys(\beta) \left( 1 + \int_0^\beta \dd\beta' \int_0^{\beta'} \dd\beta'' \sum_{i = x, y, z} \sum_{j = x, y, z} S_i(\beta') S_j(\beta'') \, \tr_\bath \left[ \tau_\bath B_i(\beta') B_j(\beta'') \right] \right).
\end{equation}
where $\tau_\sys(\beta)$ is the Gibbs state for the system, $\tau_\sys(\beta) = e^{- \beta H_\sys} / \mathcal{Z}_\sys^{(0)}$.
\begin{equation}
    \tau_\sys = \frac{e^{- \beta H_\sys}}{\mathcal{Z}_\sys^{(0)}} = \frac{e^{\beta \larmor S_z}}{\tr_\sys \left[ e^{\beta \larmor S_z} \right]},
\end{equation}
i.e., the external field $\bm{B}_\ext$ is taken to point in the $z$-direction. Appearing in the above expression for $\tilde{\rho}_S^{(2)}$ is the bath correlation function albeit with respect to inverse temperature, $t \to -i \beta$, $G_{ij}(\beta, \beta'-\beta'')$, given by
\begin{align}
    G_{ij}(\beta,\beta'-\beta'') &= \tr_\bath \left[ \tau_\bath B_i(\beta') B_j(\beta'') \right] \notag \\
    &= \hbar \int_0^\infty \dd\omega \frac{c_\omega^2}{2\omega} \left( \langle a_{\omega, i} a^\dagger_{\omega, j} \rangle e^{-(\beta' - \beta'') \hbar \omega} + \langle a_{\omega, i}^\dagger a_{\omega, j} \rangle e^{(\beta' - \beta'') \hbar \omega} \right),
\end{align}
where $\langle \ldots \rangle = \tr_\bath \left[ \tau_\bath \ldots \right]$. For the product thermal state and for independent baths
\begin{equation}
    \langle a_{\omega, i} a^\dagger_{\omega, j} \rangle_{\mathrm{th}} = (n_\beta(\omega) + 1) \delta_{ij}, \quad 
    \langle a_{\omega, i}^\dagger a_{\omega, j} \rangle_\mathrm{th} = n_\beta(\omega) \delta_{ij}, \quad
    n_\beta(\omega) = 1/(e^{\beta \hbar \omega} - 1).
\end{equation}
Thus we can write
\begin{equation}
    G_{ij}(\beta, \beta'-\beta'') = \delta_{ij} G(\beta, \beta'-\beta'')
\end{equation}
with
\begin{equation}
    G(\beta, \beta') = \hbar \int_0^\infty \dd\omega \, J(\omega) \left( (n_\beta(\omega) + 1) e^{- \beta' \hbar \omega} + n_\beta(\omega) e^{\beta' \hbar \omega} \right),
\end{equation}
where we have identified the spectral density function $J(\omega)=c_\omega^2/(2\omega)$. Thus
\begin{equation}
    \tilde{\rho}_\sys^{(2)} = \mathcal{Z}_\sys^{(0)} \mathcal{Z}_\bath \, \tau_\sys(\beta) \left(1 + \int_0^\beta \dd\beta' \int_0^{\beta'} \dd\beta'' \sum_{i = x, y, z} S_i(\beta') S_i(\beta'') \, G(\beta, \beta'-\beta'') \right),
\end{equation}
and hence
\begin{equation}
    \mathcal{Z}_\sysbath^{(2)} = \tr_\sys \left[ \tilde{\rho}_\sys^{(2)} \right] = \mathcal{Z}_\sys^{(0)} \mathcal{Z}_\bath \left( 1 + \int_0^\beta \dd\beta' \int_0^{\beta'} \dd\beta'' \sum_{i = x, y, z} \tr_\sys \left[ \tau_\sys(\beta) S_i(\beta') S_i(\beta'-\beta'') \right] \, G(\beta, \beta'') \right).
\end{equation}
The system correlation function $\sum_i \tr_\sys \left[ \tau_\sys(\beta) S_i(\beta') S_i(\beta'-\beta'') \right]$ can be readily evaluated using $H_\sys = -\larmor S_z$ and $\tau_\sys(\beta) S_\pm = S_\pm \tau_\sys(\beta) e^{\mp \beta \hbar \larmor}$. Further simplification follows by making use of the spin operator commutation rules. After carrying out the double integral and taking the trace over the system, we find that the reduced partition function for the system will then be, with $\mathcal{Z}_\sys^{(2)} = \tr_\sys \left[ \tilde{\rho}_\sys^{(2)} \right] / \mathcal{Z}_\bath$, given by
\begin{equation} \label{eq:weak_partition_1}
    \mathcal{Z}_\sys^{(2)} = \tr \left[e^{\beta \larmor S_z} \right] + \beta \tr \left[e^{\beta \larmor S_z} S^2 \right] I_1(\larmor) + \beta \tr \left[e^{\beta \larmor S_z} S_z^2\right] I_2(\larmor) - \beta \tr \left[e^{\beta \larmor S_z} S_z \right] I_3(\beta, \larmor),
\end{equation}
with
\begin{align}
    I_1(\larmor) &= \int_0^\infty \dd\omega \, J(\omega) \frac{\omega}{\omega^2 - \larmor^2}, \label{eq:I1} \\
    I_2(\larmor) &= \int_0^\infty \dd\omega \, J(\omega) \left( \frac{1}{\omega} - \frac{\omega}{\omega^2 - \larmor^2} \right), \label{eq:I2}
    \\
    I_3(\beta, \larmor) &= \hbar \int_0^\infty \dd\omega \, J(\omega) \coth \left( \frac{\beta \hbar \omega}{2} \right) \frac{\larmor}{\omega^2 - \larmor^2}
    \label{eq:I3}.
\end{align}
We can use the partition function to calculate the expectation value of $S_z$ as follows
\begin{equation} \label{eq:weak_expectation_1}
    \langle S_z \rangle = \frac{1}{\beta} \frac{1}{\mathcal{Z}_\sys^{(2)}} \frac{\partial \mathcal{Z}_\sys^{(2)}}{\partial \larmor}.
\end{equation}
For the case of spin-1/2 considered here, $S_0 = \hbar/2$, we can easily evaluate the traces in \eqref{eq:weak_partition_1} as
\begin{align}
    \mathcal{Z}_\sys^{(2)} &= 2 \cosh (\beta \larmor S_0) \left( 1 + \beta S_0(S_0 + \hbar) I_1(\larmor) + \beta S_0^2 I_2(\larmor) \right) - 2 \beta S_0 \sinh (\beta \larmor S_0) I_3(\beta, \larmor), \label{eq:weak_partition_2} \\
    &= 2 \cosh (\beta \larmor S_0) \left( 1 + \beta S_0(S_0 + \hbar) I_1(\larmor) + \beta S_0^2 I_2(\larmor) -\beta S_0 \tanh (\beta \larmor S_0) I_3 (\beta, \larmor) \right). \label{eq:weak_partition_3}
\end{align}
We now take the derivative of \eqref{eq:weak_partition_3} with respect to $\larmor$
\begin{multline} \label{eq:weak_partition_derivative}
    \frac{\partial \mathcal{Z}}{\partial \larmor} = \beta S_0 \tanh (\beta \larmor S_0) \mathcal{Z} + 2 \cosh (\beta \larmor S_0) \bigg( \beta S_0(S_0 + \hbar) I_1'(\larmor) + \beta S_0^2 I_2'(\larmor) \\ - \beta S_0 \tanh(\beta \larmor S_0) I_3'(\beta, \larmor) - \beta^2 S_0^2 \sech^2(\beta \larmor S_0) I_3(\beta, \larmor) \bigg).
\end{multline}
Substituting both \eqref{eq:weak_partition_3} and \eqref{eq:weak_partition_derivative} into \eqref{eq:weak_expectation_1} gives
\begin{multline} \label{eq:weak_expectation_2}
    \langle S_z \rangle = S_0 \tanh (\beta \larmor S_0) \\ + \frac{ S_0(S_0 + \hbar) I_1'(\larmor) + S_0^2 I_2'(\larmor) - S_0 \tanh(\beta \larmor S_0) I_3'(\beta, \larmor) - \beta S_0^2  \sech^2(\beta \larmor S_0) I_3(\beta, \larmor) }{1 + \beta S_0(S_0 + \hbar) I_1(\larmor) + \beta S_0^2 I_2(\larmor) - \beta S_0 \tanh(\beta \larmor S_0) I_3(\beta, \larmor) }.
\end{multline}
Working to the lowest order in the coupling, i.e. to the first order in $\alpha$, leaves us with
\begin{multline} \label{eq:weak_expectation_3}
    \langle S_z \rangle = S_0 \tanh(\beta \larmor S_0) + S_0(S_0 + \hbar) I_1'(\larmor) + S_0^2 I_2'(\larmor) - S_0 \tanh(\beta \larmor S_0) I_3'(\beta, \larmor) \\ - \beta S_0^2 \sech^2(\beta \larmor S_0) I_3(\beta, \larmor).
\end{multline}
In the $T \to 0$~K limit, $\beta \to \infty$, such that
\begin{equation} \label{eq:weak_expectation_zero_T_1}
    \langle S_z \rangle_{T=0} = S_0 \left[ 1 + \left( \hbar I_1'(\larmor) - I_3'^\infty(\larmor) \right) + S_0 \left( I_1'(\larmor) + I_2'(\larmor) \right) \right],
\end{equation}
where
\begin{equation} \label{eq:I3_inf}
    I_3^\infty(\larmor) = \hbar \int_0^\infty \dd\omega \, J(\omega) \frac{\larmor}{\omega^2 - \larmor^2}.
\end{equation}
\noindent
Explicitly evaluating the sum of integrals in \eqref{eq:weak_expectation_zero_T_1} allows us to express the expectation value at $T=0$~K as
\begin{equation}
    \langle S_z \rangle_{T=0} = S_0 \left[ 1 - \hbar \int_0^\infty \dd\omega \, \frac{J(\omega)}{(\omega + \larmor)^2}\right].
\end{equation}

\end{document}